\documentclass[12pt]{article}
\usepackage{amsfonts,epsfig,graphics}

\title{On Domain-wall/QFT dualities in various dimensions}
\author{Eric Bergshoeff and Rein Halbersma,\\
{\it Institute for Theoretical Physics, University of Groningen,}\\
{\it Nijenborgh 4, 9747 AG Groningen, The Netherlands}}

\begin{document}

\maketitle

\begin{abstract}

We investigate domain-wall/quantum field theory correspondences in various
dimensions. We give particular emphasis to the special case
of the quantum mechanics of 0--branes.

\end{abstract}

\section*{Introduction}
Anti-de Sitter (AdS) gravity has attracted much attention due to the
conjectured correspondence to a conformal field theory
(CFT) on the boundary of the AdS spacetime  \cite{malda} 
leading to the so-called AdS/CFT correspondence.
(for a review, see \cite{review}). The AdS/CFT correspondence has been
extended to a DW/QFT
correspondence \cite{IMSY, BST} for D$p$--branes in ten dimensions.
In this talk we extend the discussion of \cite{IMSY, BST}
to general two-block $p$--branes in various dimensions. 

In Section 1 we 
first discuss some general facts on Domain-Wall and anti-de Sitter spacetimes.
In Section 2 we calculate
the near-horizon geometries of a generic $p$--brane. 
In Section 3 we discuss the field theory limit for the general case.
The formulae we give in this Section are applied
to the special case of 0--branes we discuss in the Section 4.
In this final Section  we discuss the quantum mechanics of 0--branes, or
extreme black holes, in various dimensions.

\section{Domain-Walls and anti-de Sitter spacetimes}

Domain-wall (DW) spaces, i.e.~spaces of co-dimension 1,
occur as solutions to the equations of motion of a (super-)gravity action 
with a dilaton scalar $\phi$ and a 
(D-1)--form gauge potential (for a
review, see \cite{Cv1}). They are $p$--branes with 
worldvolume dimension $p+1$ which is one less than the dimension $D$ of 
their target spacetime, i.e.~$D=p+2$. 
The part of the supergravity action needed to describe the DW solution 
is given by (we use the Einstein frame and mostly plus signature):
\begin{equation}
\label{einstcosm}
S (D,b) =  \int {\rm d}^{D} x \; {\sqrt{-g}\over 2\kappa_D^2}
 \left[ - R - 
{\textstyle{4 \over D-2}} 
(\partial \phi)^2 
- {\textstyle {1\over 2D!}}\,
g_s^{2k} \left ({e^{\phi}\over g_s}\right )^b\, F_D^2
 \right]\, ,
\end{equation}
where $\kappa_D$ is the gravitational coupling constant with 
(ignoring constants)
\begin{equation}
\label{kappaD}
\kappa_D^2 \sim  \ell_s^{D-2} g_s^2\, ,
\end{equation}
 and
$b$ is the dilaton coupling parameter. We also introduced a parameter $k$
 defined by
\begin{equation}
\label{park}
k\,(D,b) = {b\over 2} + 2 {D-1\over D-2}\, .
\end{equation}
Solving the equations of motions following from (\ref{einstcosm}) one finds:
\begin{eqnarray}
\label{domainwall}
{\rm d}s^2 &=& H^{- {4 \varepsilon \over (D-2) \Delta_{\rm DW}}}\, 
{\rm d}x_{D-1}^2 +
H^{{-4 \varepsilon (D-1) \over (D-2) \Delta_{\rm DW}}  
-2 (\varepsilon +1)}\, {\rm d}y^2\, , \nonumber \\ 
e^{\phi} &=& g_s\, H^{-(D-2) b \varepsilon \over 4 \Delta_{\rm DW}}\, ,\\
g_s^{k}F_{01 \ldots D-2\,y}&=& \pm\, \sqrt{{4\over \Delta_{\rm DW}}}
\partial_y H^\varepsilon\, , \nonumber 
\end{eqnarray}
with  $\varepsilon$ a parameter and where $g_s$ and 
$\Delta_{\rm DW}$ (which is invariant 
under dimensional reduction) are defined by
\begin{eqnarray}
\label{deldw}
g_s &=& e^{\phi (H=1)}\, ,\nonumber\\
\Delta_{\rm DW}\,(D,b) &=& {\textstyle{1\over 8}}(D-2)\, b^2 - 2{D-1 \over D-2}
\, .
\end{eqnarray}
The function $H$ is harmonic on the 1-dimensional transverse space with 
coordinate $y$:
\begin{eqnarray}
H &=& c + Q_1 y\, ,\hskip 1truecm y>0\, , \nonumber\\
H &=& c+ Q_2 y\, ,\hskip 1truecm y<0\, ,
\end{eqnarray}
with $c, Q_1, Q_2$ constant. The domain-wall is 
positioned at the discontinuity $y=0$.
Different choices of $\varepsilon$ correspond to different choices
of coordinates and lead to different expressions for the 
metric \cite{M9}. For instance, one can choose $\varepsilon$ such that the 
powers of the harmonic function become equal or opposite.

Shifting the position of the domain 
wall to infinity, e.g. $y \rightarrow +\infty$
allows us to discard the constant $c$ in the harmonic 
function and, furthermore there is only one side of the domain wall. 
We can eliminate $\varepsilon$ if we define a mass parameter $m$ 
by :
\begin{equation}
Q \varepsilon = m\, ,
\end{equation}
with $Q = Q_1$.
Making a co-ordinate transformation and by going to the so-called
``dual frame'' metric $g_{*}$, indicated by a subscript star, 
\begin{equation}
\label{reg}
Qy =e^{-Q\lambda},  \hskip 2truecm  g_{*}=e^{-b \phi} g_{E} ,
\end{equation}
we find for the solution (\ref{domainwall})
\begin{equation}
\label{dwr}
{\rm d}s^2_{*} =  e^{-2m \lambda ({2+\Delta_{\rm DW} \over \Delta_{\rm DW}})} 
{\rm d}x_{D-1}^2 + {\rm d}\lambda^2 , \hskip 1truecm
\phi = \lambda {(D-2) b m \over 4 \Delta_{\rm DW}}.
\end{equation}
In the dual frame (\ref{reg}) the domain-wall solution (\ref{dwr}) describes 
an ${\rm AdS}_{D}$ spacetime with a linear dilaton, the latter breaking 
the full conformal structure of the AdS spacetime. This observation will 
be the key to generalizing the AdS/CFT duality to a DW/QFT duality.

\section{Near-horizon Geometries of $p$--branes}

Our starting point is the $D$-dimensional action
\begin{equation}
\label{generalaction}
S(D,a,p) = \int {\rm d}^D x \; {\sqrt{-g}\over 2\kappa_D^2}
 \Big[ -R - {\textstyle{4 \over D-2}} (\partial \phi)^2 
- {g_s^{2k}\over 2(d + 1)!}\left ( {e^\phi\over g_s}\right )^{a}
  F_{ d +1}^2 \Big]\, ,
\end{equation}
which contains three independent parameters: the target spacetime dimension 
$D$, the dilaton coupling parameter $a$ and a parameter $p$ specifying the 
rank $D-p-2$ of the  field strength $F$. The parameter $k$ is a generalization
of (\ref{park}) and is given by
\begin{equation}
k(D,a,p) = {a\over 2} + 2{p+1\over D-2}\, .
\end{equation}
We have furthermore introduced two 
useful dependent parameters $d$ and $\tilde d$ which are defined by 
\begin{equation}
\left\{
\begin{array}{rcl}
d &=& p+1 \ \ \hskip .9truecm  {\rm dimension\ of\ the\ worldvolume}\, , \\
\tilde d &=& D- p-3  \ \  {\rm dimension\ of\ the\ dual\ brane\
worldvolume}\, .
\end{array}
\right. 
\end{equation}
We next consider the following class of diagonal ``two-block'' $p$--brane 
solutions (using the Einstein frame)\footnote{For later convenience, we give 
the solution in terms of the {\it magnetic} 
potential of rank $D-p-3$. 
The p-brane solution is {\it electrically} charged with respect
to the p+1--form potential.}:
\begin{eqnarray}
\label{2block}
{\rm d}s^2 &=& H^{- {4\tilde d \over (D-2) \Delta}} {\rm d}x_{d}^2 +
H^{4d \over (D-2) \Delta} {\rm d}x_{\tilde d + 2}^2\, ,\nonumber\\
e^{\phi} &=& g_s\,H^{(D-2) a \over 4 \Delta}\, , \\
g_s^{(2-k)} {\tilde F}_{m_1\cdots m_{\tilde d + 1}} &=& 
\pm \sqrt {{4\over \Delta}} 
\epsilon_{m_1\cdots m_{\tilde d +1}m}\partial_m H\nonumber\, ,
\end{eqnarray}
where
\begin{equation}
g_s^{(2-k)}\, \tilde F = \left ({e^\phi\over g_s}
\right )^a g_s^{k}\, {}^* F\, .
\end{equation}
We use a constant, i.e.~metric independent, Levi-Civita tensor. Furthermore,
$g_s = e^{\phi(H=1)}$ and $\Delta$ is a 
generalization of the $\Delta_{\rm DW}$ in the previous section defined by
\cite{Pope}
\begin{equation}
\label{delta}
\Delta = {\textstyle{1\over 8}}(D-2) a^2  + { 2d \tilde d \over D-2}\, ,
\end{equation}
which is invariant under reductions and oxidations (in the Einstein frame). 
The function $H$ is harmonic over the $\tilde d + 2$ transverse coordinates 
and, assuming that 
\begin{equation}
\label{transconstr}
{\tilde d} \ne -2, 0\, ,
\end{equation}
(i.e. no constant or logarithmic harmonic) this harmonic function is given by
\begin{equation}
H = 1 + \Big({r_0 \over r}\Big)^{\tilde d}\, .
\end{equation}
Here $r_0$ is an integration constant with the dimensions of length. It 
is related to the mass and charge of the $p$--brane a follows. The mass
$\tau_p$ per unit p--volume is given by the ADM--formula:
\begin{eqnarray}
\label{taup}
\tau_p &=& {1\over 2\kappa_D^2}\int_{\partial M^{D-p-1}} d^{D-p-2}
\Sigma^m\left (
\partial^nh_{mn} - \partial_mh^b{}_b\right )\nonumber\\
&=& {2(D-p-3)\over \Delta \kappa_D^2}\, r_0^{D-p-3}\,\Omega_{D-p-2}\, .
\end{eqnarray}
On the other hand, the charge $\mu_p$ per unit p-volume is given, in terms
of the same integration constant $r_0$, by
the Gauss-law formula
\begin{eqnarray}
\mu_p &=& {1\over 2\kappa_D^2} \int (d^{D-p-2}\Sigma)^{m_1\cdots m_{D-p-2}}
g_s^{(2-k)} {\tilde F}_{m_1\cdots m_{D-p-2}}\nonumber\\
&=&\pm \sqrt {{\Delta \over 4}} \tau_p\, .
\end{eqnarray}
Hence, the p--brane solution satifies the BPS bound
\begin{equation}
\tau_p = \sqrt {{4\over \Delta}}\,|\mu_p|\, .
\end{equation}
To derive an expression for $r_0$ in terms of the string parameters
$g_s$ and $\ell_s$, which fixes the scaling of $H$,
one must add a source term to the supergravity
bulk action. Using  the no-force condition\footnote{Alternatively, one can 
use a scaling argument, see Appendix B of \cite{us}.}
 and the fact that in the string
frame the electric (p+1)--form potential $C_{p+1}$ is proportional to 
$g_s^{-k}$,
as follows from the action (\ref{generalaction}), we find that 
\begin{equation}
\tau_p \sim {1\over \ell_s^{p+1} g_s^k}\, .
\end{equation}
Comparing with (\ref{taup}) and using (\ref{kappaD}) we deduce that,
for a single brane,
\begin{equation}
\left ( {r_0\over \ell_s}\right )^{\tilde d} \sim g_s^{2-k}\, .
\end{equation}

The two-block solutions (\ref{2block}) include the (supersymmetric) 
domain-wall spaces of the previous section. They correspond to the case 
$\tilde d=-1$, $\varepsilon=-1$ and $r_0 = 1/m$. The solutions also include
 the known branes in ten and eleven dimensions (M2, M5, D$p$, F1, NS5 etc.)
 as well as branes in lower dimensions. If the branes under consideration 
preserve any supersymmetries we can set \cite{Pope}
\begin{equation}
\label{susydelta}
\Delta = {4 \over n}\, ,
\end{equation}
where generically $32/{2^n}$ is the number of unbroken supersymmetries.

We now consider the limit for which the constant part in the harmonic 
function is negligible. As in the previous section we make a co-ordinate
 transformation and go to the dual frame
\begin{equation}
\label{dualframe}
\left({r_0 \over r}\right) = e^{-\lambda/r_0}\, \quad g_{*} = e^{({a \over 
\tilde d})\phi}  g_E\, .
\end{equation}
After these manipulations we can write the near-horizon metric as
\begin{equation}
\label{nhgeometry}
{\rm d}s^2_{*} = e^{-2(1 - {2 \tilde d \over \Delta }) \lambda/r_0 } 
{\rm d}x_{d}^2 + {\rm d} \lambda^2 + r_0^2 {\rm d} \Omega_{\tilde d + 1}^2
\, , \qquad 
\phi = -\lambda {(D-2) a \tilde d \over 4 \Delta r_0} \, ,
\end{equation}
which has an $AdS_{d+1} \otimes S^{\tilde d + 1}$  geometry and a linear 
dilaton.
 
Reducing over the $\tilde d + 1$ angular variables of the sphere we 
end up with a gauged supergravity in $d+1$ dimensions of the form 
(\ref{einstcosm}) supporting a domain-wall solution of the form 
(\ref{domainwall}). The precise relation between the parameters of the 
action (\ref{einstcosm}) and its solution (\ref{domainwall}) in terms of 
those of the action (\ref{generalaction}) and its solution (\ref{nhgeometry})
 can be found in section 2.3 of \cite{us}.

Summarizing, in this Section we showed that in the dual frame, defined by
(\ref{dualframe}), all $p$--branes solutions (\ref{2block}) have a near 
horizon geometry ${\rm DW}_{d+1} \otimes S^{\tilde d +1}$. The 
domain-wall metric has all the isometries of an AdS space. These isometries 
are broken in the full background because of the presence of a linear
 dilaton.

\section{The field theory limit}

In this section we will set up the framework for the DW/QFT duality similar 
to the analysis of \cite{BST} but for arbitrary dimensions. The analysis
is similar to the AdS/CFT duality. There one finds (in the low-energy limit)
 that supergravity in the near-horizon geometry of a large number of 
non-dilatonic branes is dual to the conformal field theory living on these 
branes (which are located at the boundary of the near-horizon geometry). 
This so-called holographic principle lies at the heart of the AdS/CFT 
conjecture.

As we showed in the previous sections, the near-horizon geometry of a general 
$p$-brane in the dual-frame is equivalent to that of a non-dilatonic 
$p$-brane. It is therefore natural to assume that the duality might be 
extended. The presence of the dilaton makes the AdS background into a 
domain-wall background and the conformal field theory into a general 
quantum field theory, hence the name DW/QFT duality. In the following 
we will only consider D-branes and their intersections in lower dimensions.

One starts with a D$p$-brane configuration in string theory and takes the 
low-energy limit in which only the massless modes survive and in which the 
theories in the bulk and on the brane decouple.
Next, one introduces a new energy-scale and a dimensionless coupling 
constant which are the free parameters of the decoupled theories. 
The energy-scale can be given in several ways:
\begin{figure}[h]
\label{probes}
\begin{center}
\scalebox{.45}{\input{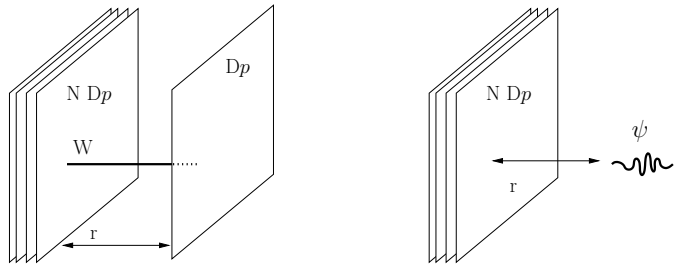}}
\caption{Probing a stack of $N$ D$p$-branes can either be done by another
D$p$-brane (left) or by a supergravity field $\psi$ (right)
} 
\end{center}   
\end{figure}
If one probes the stacked branes by another D$p$-brane, a natural 
energy-scale is
 given by the mass of  the endpoint of a stretched string, which acts like
a W--boson (see Figure 1)
\begin{equation}
E_{W} = U = {r \over l_s^2}\, .
\end{equation}
This was the approach taken by \cite{IMSY}, all branes have the same 
energy-scale but the disadvantage is that the near-horizon geometry when 
written in the $U$ co-ordinates does not take the same form for all branes.
One can also probe the branes by a supergravity field $\psi$ 
(suppressing possible quantum numbers). By solving  the wave 
equation for this field \cite{PP} one finds the following energy scale
\begin{equation}
E_{\psi} = u = {r^{\beta} \over r_0^{\beta + 1}}, \quad \beta = 
{2 \tilde d \over \Delta} - 1, \quad r_0 = l_s (g_s^{2-k}
 N)^{1 \over \tilde d}\, .
\end{equation}
In \cite{SW} it was shown that this energy-scale corresponds to the 
so-called holographic energy-scale used in entropy calculations. 
Although the energy depends on the parameters of the brane solution, 
one finds that the near-horizon geometry (\ref{dwr}) in these holographic 
$u$ co-ordinates is particularly simple:
\begin{equation}
\label{limitsol}
{\rm d}s_{*}^2 = r_0^2 \left[ \mathcal{R}^2 \left\{ u^2 {\rm d}x^2_{d} + 
\left({{\rm d}u \over u}\right)^2 \right\} + {\rm d}\Omega_{\tilde d + 1}^2 
\right]\, .
\end{equation}
For all branes this metric has the form $AdS(\mathcal{R} r_0)_{d+1} \otimes
S^{\tilde d+1}(r_0)$ 
differing only in the (relative) radii of the sphere and 
anti-de-Sitter space.

The quantum field theory on the brane is in general not explicitly known, 
but dimensional analysis  enables us to extract some information. If we 
assume that the theory is given by some $q$-form potential on the $d=p+1$ 
dimensional worldvolume
\begin{equation}
S_{wv} = \tau_p \int {\rm d}^{p+1} \; \xi\, {\rm Tr}\, F_{q+1}^2\, ,
\end{equation}
then we have for the coupling constant scaling behavior
\begin{equation}
g_{ft} \sim g_s^k\,  l_s^{\alpha} \hskip 1truecm {\rm with}\hskip 1truecm
 \alpha = p - 2 q - 1.
\end{equation}
For 10-dimensional D$p$-branes we have a vector-multiplet (i.e.~$q=1$) on the 
worldvolume reproducing the familiar $\alpha = p-3$ scaling behavior for the 
Yang-Mills theory on the D$p$-brane worldvolume. 
The effective dimensionless coupling 
constant $\lambda$ is given by
\begin{equation}
\lambda_W = g_{ft} N E_W^{\alpha}\, ,\hskip 2truecm
\lambda_\psi = g_{ft} N E_\psi^{\alpha}\, . 
\end{equation}
Since we have two different probes, we can in principle construct two 
different dimensionless couplings, constructed by either $E_{W}$ or 
$E_{\psi}$. Since both probes have a sensible interpretation, it must be 
possible to use them both, independently of the low-energy limit. This 
leads to the following constraint on the scaling behavior in terms of the 
brane-solution parameters \cite{us}:
\begin{equation}
\label{scaleconstr}
\alpha = \Delta - \tilde d = a {D-2 \over 4}\, .
\end{equation}
Combining this with the definitions of the probe-scales $E_W$ and 
$E_{\psi}$ we find the following relations between them and their 
corresponding dimensionless couplings $\lambda_{W}$ and $\lambda_{\psi}$
\begin{equation}
\label{proberel}
{E_{W} \over E_{\psi}} = \lambda_{W}^{2 \over \Delta}, \hskip 1.5truecm
\lambda_{W}^{\beta} =  \lambda_{\psi}\, .
\end{equation}
The decoupling limit is now given by
\[
\begin{array}{ll}
\mbox{take} & {E/ E_s} \rightarrow 0, \hskip 2truecm E_s = \ell_s^{-1}\, ,\\
\mbox{at fixed} &  
\left\{\begin{array}{ll}
\lambda_W = g_{\rm ft} N E_W^{\alpha}\quad {\rm and}\quad {E/ E_{\rm W}}
\hskip 1truecm 
{\rm or,\, equivalently,\, see\, (\ref{proberel}),}\\
\lambda_\psi = g_{\rm ft} N E_\psi^{\alpha}\quad\ \ {\rm and}\quad 
{E / E_{\rm \psi}}
\end{array}\right..
\end{array}
\]
The dimensionless coupling made from Newton's constant in this limit becomes
\begin{equation}
G_{N} = g_s^2 \left({E \over E_s}\right)^{D-2} = 
\left({\lambda \over N}\right)^2 \left({E \over E_s}\right)^{D-2 - 2 \alpha} 
\end{equation}
so that if $\alpha$ becomes too large one has to take $N$ to infinity to
 ensure the decoupling of gravity.

In the AdS/CFT correspondence one has in the field theory side the parameters
 $(\lambda, {1 \over N})$ which correspond to the curvature and string 
coupling on the supergravity side. They generalize to the effective 
tension of a string in the dual frame and the dilaton, respectively:
\begin{equation}
\tau_{*} l_s^2 = \lambda_{\psi}^{2 \over \Delta \beta}, \quad
e^{\phi} = {\tau_{*}^{\tilde d \over 2} \over N}.
\end{equation}
We now have the following ranges of validity on the two sides of the duality
\begin{itemize}
\item perturbative SYM: 
\[
\lambda_{W} \ll 1 \Rightarrow
\left\{ 
\begin{array}{lll}
\alpha < 0: E_{W} \rightarrow \infty: & \mbox{UV-free,}\\
\alpha > 0: E_{W} \rightarrow 0: & \mbox{IR-free.}
\end{array}
\right.
\]
\item classical SUGRA: 
\[
\left\{ 
\begin{array}{lll}
\tau_{*} l_s^2 & \gg 1: & \mbox{no stringy corrections,}\\
e^\phi & \ll 1: & \mbox{no quantum corrections.}
\end{array}
\right.
\]
\end{itemize}
Using the above formulae one can easily see that classical supergravity 
describes strongly coupled, large N field theory. However, the conformal 
invariance which in the AdS/CFT duality facilitates computations in the 
strongly coupled field theory is now broken so that any direct check of a 
DW/QFT duality is ruled out. For specific examples, we refer to \cite{us}.

As long as $\beta$ is positive, the supergravity and field theory are valid in 
different regimes (i.e.~$\lambda_{\psi} \gg 1$ and $\lambda_{W} \ll 1$) and 
one can find consistent phase diagrams as in \cite{IMSY}. However, for
negative 
$\beta$, e.g.~in case of the D6-brane,
 this seems no longer true at first sight: $\lambda_{\psi} \ll 1$ and 
$\lambda_{W} \ll 1$. 

As one can see by using the relation (\ref{proberel}) between the two 
couplings, this corresponds to $\lambda_{W} \gg 1$ and $\lambda_{W} \ll 1$. 
This matches nicely with the explanation of \cite{PP}, namely that the 
low-energy Hilbert space of the field theory has two separate sectors: 
one describing nearby (in U co-ordinates) brane probes and one describing 
supergravity probes far away (also in U co-ordinates) from the brane.

Finally we note that when the dilaton becomes larger one can 
perform an S-duality transformation. If the curvature is small with 
respect to the S-dual string scale, then one can still trust the 
supergravity approximation.

\section{Dynamics of 0--branes}

In this Section we consider the special case of 0--branes in various 
dimensions. The (bosonic) Lagrangian for a particle with mass $m$ and 
charge $q$, moving in the string frame near-horizon background of $N$ 
stacked 0--branes reads
\begin{equation}
{\cal L}= m \,  e^{-\phi} \sqrt{| \dot{x}^{\mu} \dot{x}^{\nu} 
g_{\mu\nu}^{S} | } +q A_{\mu} \dot{x}^{\mu} \, , 
\end{equation}
where the dot represents derivatives with respect to the worldline time. 
The D-dimensional 0--brane solution in the dual frame $g_{\mu \nu}^{*}$ is 
given by the expression (\ref{limitsol}), taken for $p=0$. Introducing the 
canonical momentum $P_{\mu}={\partial {\cal L} \over \partial \dot{x}^{\mu}}$
 one can write down the mass-shell constraint in the string frame for the
 probe particle as
\begin{equation}
\label{mass-shell}
(P_{\mu} -qA_{\mu})(P_{\nu} -qA_{\nu})\, g^{\mu \nu}_{S} = m^2 \, e^{-2\phi} 
\, .
\end{equation}

We would like to solve this equation for $P_t=-{\cal H}$. For this purpose,
we transform the 
mass-shell equation (\ref{mass-shell}) to the dual frame and substitute the 
metric and gauge field of the solution (\ref{limitsol}). For the gauge field 
(being in the temporal gauge) we can write: 
\begin{equation}
{A_t \over u} = r_0 e^{-{D-4 \over D-3} \phi} \equiv M\, ,
\end{equation}
and we find for the Hamiltonian
\begin{eqnarray*}
{\cal H}&=&\left({u \over {\cal R}}\right)^3 {P_u^2 \over 2 f} + {u \over 
{\cal R}} {g \over 2 f}\, , \\
f &=& {1 \over 2} \left(q M + \sqrt{(m M)^2 + ({u P_u \over {\cal R}})^2 + 
\vec{L}^2 } \; \right)\, ,\\
g &=& \vec{L}^2 + (m^2 - q^2) M^2\, . 
\end{eqnarray*}
We can write this Hamiltonian in a rather suggestive form by making the 
following transformation
\begin{equation}
{u \over {\cal R}} = {(2 {\cal R})^2 M \over x^2}\, .
\end{equation}
after which the Hamiltonian takes on the following form
\begin{eqnarray*}
{\cal H}&=& {P_x^2 \over 2 f} + {g \over 2 f x^2}\, , \\
f &=& {1 \over 2} \left(q + \sqrt{m^2 + \left({x P_x \over 2 
{\cal R} M}\right)^2 + \left({\vec{L} \over M}\right)^2} \; \right)\, ,\\
g &=& 4 \left(\vec{L}^2 + (m^2 - q^2) M^2 \right)\, .
\end{eqnarray*}
The dynamics of the system is now generated by the Poisson brackets with 
respect to the canonical co-ordinate and momentum $(x,P_x)$. The transition 
to quantum mechanics is  the usual one. 
Defining two other operators
\begin{equation}
{\cal D} = {1 \over 2} x P_x, \quad {\cal K} = -{1 \over 2} f x^2,
\end{equation}
we find that the system possesses a classical $SL(2,\mathbb{R})$-symmetry
\begin{equation}
\{{\cal D}, {\cal H}\} = {\cal H} \quad
\{{\cal D}, {\cal K}\} = -{\cal K} \quad
\{{\cal H}, {\cal K}\} = 2{\cal D}\, .
\end{equation}
This is only true when the effective coupling constant $\lambda_{\psi}$ of 
the model under consideration is fixed under scale transformations. Whenever 
we have a non-trivial dilaton the model will not be conformally invariant by 
itself. Only when introducing a transformation to keep $\lambda_{\psi}$ fixed 
under conformal transformations, will the model be invariant under what are 
called generalized conformal transformations \cite{jeyo}.

Taking the limit $M m \rightarrow \infty$ and at the same time 
$M (m-q) \rightarrow 0$ \cite{derix} we find a ``non-relativistic'' 
Hamiltonian of the form:
\begin{equation}
{\cal H} = {P_x^2 \over 2 m} + {\vec L^2 \over 2 m x^2}\, .
\end{equation}
This model, as first considered in \cite{AFF}, is in fact an integrable model
 so that it might offer a concrete test of the $DW_{2}/QFT_{1}$ duality in 
this particular limit.

\section*{Acknowledgements}

The work reported here is based upon {\tt hep-th/9907006}. We thank
our collaborators Klaus Behrndt and Jan Pieter van der Schaar for numerous
discussions.
This work is supported by the European Commission TMR programme
ERBFMRX-CT96-0045, in which E.B. and R.H. are  associated 
to the University of
Utrecht.


\end{document}